\documentclass[showkeys,secnumarabic,showpacs,amsmath,amssymb,prd]{revtex4}

\begin{document}

\title{Black holes in the varying speed of light theory}

\author{Hossein Shojaie}
 \email{h-shojaie@sbu.ac.ir}

\author{Mehrdad Farhoudi}
 \email{m-farhoudi@sbu.ac.ir}

\affiliation{Department of Physics, Shahid Beheshti University,\\
Evin, Tehran 1983963113, Iran}

\begin{abstract}
We consider the effect of the \emph{Varying Speed of Light}
theory on non-rotating black holes. We show that in any
varying-$c$ theory, the Schwarzschild solution is neither static
nor stationary. For a no-charged black hole, the singularity
in the Schwarzschild horizon cannot be removed by coordinate
transformation. Hence, no matter can enter the horizon, and the
interior part of the black hole is separated from the rest of the
Universe. If $\dot{c}<0$, then the size of the Schwarzschild
radius increases with time. The higher value of the speed of light
in the very early Universe may have caused a large reduction in the probability of the creation
of the primordial black holes and their population.The same analogy is also considered for the charged black
holes.
\end{abstract}



\maketitle

\section{Introduction}

Although the \emph{Standard Big Bang} (SBB) model of the Universe
provides a successful model and does not conflict with astronomical
observations, several preassumptions in this model, which were
introduced as initial values rather than derived aspects, can be
considered as its weakness. According to the SBB model the flatness,
horizon, primordial seeds of galaxies, magnetic monopole, and
cosmological constant ($\Lambda$) problems cannot be derived from
the theory~\cite{r20}.

Efforts to derive the above problems as results of the theory
have mainly introduced the \emph{inflationary} scenario of the
Universe~\cite{r1,r2,r3,r4}. In this scenario, immediately after
the big bang, the Universe
experienced a superluminal expansion due to potentials that
convert gravitational attraction into repulsion. During the last two
decades of twentieth century, the various  models of inflation
were improved~\cite{r25} but there are still some open questions
that must be answered by these models. For instance, no
successful microscopic foundation for inflation has been
presented. At this time, no models of inflation are fully
satisfactory~\cite{r5}; hence, one can feel free to search for
other theories.

\emph{Varying Speed of Light} (VSL) models have therefore been
introduced as alternatives to the inflationary
models~\cite{r5,r6,r7,r8,r9,r27,r29}. They solve the problems of SBB model
and are also in agreement with the theories that allow the fine
structure constant, $\alpha$, to vary, for instance, theories
that try to bind high-energy physics and standard cosmology with
the idea of dimensional reduction. Since high-energy theories use
higher dimensional spaces and low-energy physics is
four-dimensional, then there must be a dimension reduction
mechanism to lower dimensions. Such a mechanism usually lets one
or more of the constants have time or space
dependencies~\cite{r21,r22}. Observations also confirm the time
variation of $\alpha$~\cite{r10}. However, any variation in
$\alpha=e^{2}/\hbar c$ can be explained as a variation in $c$,
$\hbar$ and (or) $e$ \cite{r12}. The theory in which $e$ has a time
dependency was proposed by Beckenstein~\cite{r11}. In this theory,
the vacuum, as a dielectric medium, screens the electric charge.
Although VSL theory can be transformed into a varying-$e$ theory by
a suitable choice of transformation, the dual of minimal VSL
theory is not a minimal varying-$e$ theory itself~\cite{r12}.
Also, new observations of some supernovae show that the universe
is accelerating~\cite{r16,r17,r18}, which is in agreement
with VSL theory.

Different models of the VSL have been introduced,
but none of them can be regarded as an exact mechanism for
the dynamics of $c$ as yet~\cite{r7,r9,r13,r14,r15}. For instance, in one
model, Albrecht and Magueijo~\cite{r6} postulated that light
traveled much faster in the early Universe and, due to a phase
transition at a critical time $t_{c}$, the velocity of light fell suddenly and changed to its current value. In a model proposed
by Barrow~\cite{r7}, $c$ is a smooth power-law function of
time and reaches from a very large value to today's value. In
another model~\cite{r13,r14,r15}, a phase of spontaneous breaking
of a local Lorentz-invariance generates a large increase in the
speed of light. Reformulation of this model leads to a bimetric
model. That is, there are two separate metrics, one is associated
with gravity and one with electrodynamics, which are related to
each other with a vector or gradient of a scalar field.

There has been some efforts, in the literature, to see different
consequences of the VSL models, and in this article we intend, by
a few plausible assumptions, to examine their consequence(s) for
black holes.

\section{The field equation}

As a first consequence of the VSL theory, a dynamical field should
appear somehow in gravitational theory, hence the exact form of
the Einstein equation may not be valid anymore. Actually, when
$c(x^a)$ varies, an obvious problem appears in the heart of
general relativity, the Einstein equation itself, i.e.
\begin{equation}
G_{ab}=\frac{8\pi G}{c^4}T_{ab}\label{2.1}\ \ .
\end{equation}
For the Bianchi identity implies that the covariant divergence of the left-hand side of (\ref{2.1}) must be zero, hence one gets
\begin{equation}
\nabla_a\left(\frac{8\pi G}{c^4}T^{ab}\right)=0\label{2.8}\ \ .
\end{equation}

In the classical view, when $c$ is fixed, (\ref{2.8})
is in agreement with energy-momentum conservation, but when
$c$ varies, it is not valid anymore. To solve this problem, the
following two main ideas~\cite{r29} are suggested:
\begin{enumerate}
    \item Adding other term(s) to $T^{ab}$, but it is important to note that even when one wants to derive a vacuum solution, the right-hand side of (\ref{2.1}) may not be zero.
    \item By neglecting the energy-momentum conservation, any change in $c$ can act as a source of matter creation. If it is so, any experiment that leads to violation of this conservation can be used to estimate the changing ratio of $c$. In this case, for a vacuum solution, the right-hand side is zero.
\end{enumerate}
However, there are also some works that handle the dynamics
of the VSL models, for example, see ref.~\cite{r13}.

\section{The Schwarzschild metric}

As there is not a universally accepted theory for dynamics of the VSL
models, which includes a field for varying $c$, one may find it
useful to analyze the consequences of these models even in an
elementary, simple way. For instance, we take an advantage of the
correspondence principle or the principle of minimal coupling as an
indication for further procedure in this work. As the same advantage
has almost been taken in considering a way to amend and (or) generalize the
Friedmann equations, which were written for the Einstein
equation in a preferred frame with constant $c$, to propose a varying-$c$ model in ref.~\cite{r26}.

The Schwarzschild metric is a spherically symmetric vacuum
solution to the Einstein equation that is asymptotically flat,
i.e.,
\begin{equation}
ds^2=\left(1-\frac{2m}{r}\right)c^2dt^2-\left(1-\frac{2m}{r}\right)^{-1}dr^2-r^2d\Omega^2\label{2.6}\ \ ,
\end{equation}
with $m$ as a constant of integral. The Birkhoff theorem implies
that this solution is also static. Using the weak-field limit,
which arises purely from assuming geodesic motion and the
Newtonian limit, leads to an interpretation of the constant
$m$, as a geometrized mass, to be equal to $GM/c^2$ of a
point mass $M$ situated at the origin, i.e., the Schwarzschild
metric becomes
\begin{equation}
ds^2=\left(1-\frac{2GM}{c^2r}\right)c^2dt^2-\left(1-\frac{2GM}{c^2r}\right)^{-1}dr^2-r^2d\Omega^2\label{2.2}\ \ .
\end{equation}

It is well-known that the Schwarzschild metric is not only a
vacuum solution to the Einstein equation, but it is a particular
exact vacuum solution with fixed $G$ to the Brans-Dicke theory of
gravity as well~\cite{r23}.

Now, if either $G$ or $c$ is dependent on space or time, then an
explicit calculation of (\ref{2.2}) shows that some of the
Einstein tensor components will no longer be zero and hence such a
line element obviously does not correspond to a vacuum solution of
the standard Einstein equation.

Before we proceed further, we should clear up the following
assumption as well. In a metric one usually writes $dx^0=c\:dt$, the
component of $g_{00}$ is the coefficient of $dx^0$ and not the
coefficient of $dt$ (except in the especial case of
$c=1$)\footnote[1]{Also, no difficulty would arise even if one
assumes $x^0=c\:t$, for it should be noted that one can write
$dx^0=c'\:dt$, for example, when $c$ is a polynomial function of $t$
only.}. However, if there would be a $c(x^a)$ in a component of
$g_{ab}$, its derivative(s) should appear in the Einstein tensor.
Hence, in the case of vacuum, i.e., $T_{ab}=0$, the dynamic
equation should be $G_{ab}\neq0$ in the VSL theory, as
expected in the first suggestion of the previous section. This
means that the Birkhoff theorem is no longer applicable here.

Therefore, if one looks for a spherically symmetric gravitational
field solution for a vacuum case in the VSL theory, it should be a
plausible assumption that one would expect, somehow, to have a
generalization of the Schwarzschild metric. One may justify it
better with the correspondence principle. Hence, with the aid of
the principle of minimal coupling, we assume that the simplest
generalization of this metric in the case of the VSL theory would be
as follows
\begin{equation}
ds^2=\left(1-\frac{2GM}{c(t)^2r}\right)c(t)^2dt^2-\left(1-\frac{2GM}{c(t)^2r}\right)^{-1}dr^2-r^2d\Omega^2\label{2.7}\
\ .
\end{equation}

Obviously for constant $c$, (\ref{2.7}) goes to the
standard Schwarzschild metric; and with the metric (\ref{2.7}) one
will not get the standard Einstein vacuum equation as expected.

Perhaps, it could be more accurate if one assumes $c=c(t,r)$.
However, in a preferred frame theory, as in the most of the work
appearing in the literature to date regarding the VSL models, when
cosmological solutions are considered, i.e., homogeneous and
isotropic space-time, one can assume $c$ as a function of time
only.

The metric (\ref{2.7}) is no longer static or even stationary,
this is a penalty that one must pay for this generalization. We
should emphasize that one should have expected such a result, for
a time dependence for $c$ creates extra terms on the right-hand
side of the Einstein equation. However, it is still asymptotically
flat as one can easily check.

Like the classical Schwarzschild solution, this line element in
addition to an essential singularity in the center, has a
singularity at
\begin{equation}
r_s=\frac{2GM}{c(t)^2}\ ,\label{2.3}
\end{equation}
namely, at the Schwarzschild radius.

A question arises: is the singularity at (\ref{2.3}) in the VSL theory
removable? To answer this question, one may calculate the Riemann
tensor scalar invariant as usual, but with the assumption of
varying $c$. The resultant is
\begin{eqnarray}
R_{abcd}R^{abcd}&=&12\frac{\left(\frac{r_s}{r}\right)^2}{r^4}-8\frac{\left(\frac{r_s}{r}\right)^2\ddot{c}(t)}{r^2c(t)^3(1-\frac{r_{s}}{r})^2}+16\frac{\left(\frac{r_s}{r}\right)^2\dot{c}(t)^2(1+\frac{r_{s}}{r})}{r^2c(t)^4(1-\frac{r_{s}}{r})^3}{}\nonumber\\
&&{}+4\frac{\left(\frac{r_s}{r}\right)^2\ddot{c}(t)^2}{c(t)^6(1-\frac{r_{s}}{r})^4}-32\frac{\left(\frac{r_s}{r}\right)^2\dot{c}(t)^2\ddot{c}(t)}{c(t)^7(1-\frac{r_{s}}{r})^5}{}\nonumber\\
&&{}+64\frac{\left(\frac{r_s}{r}\right)^2\dot{c}(t)^4}{c(t)^8(1-\frac{r_{s}}{r})^6}
\label{2.4}\ \ .
\end{eqnarray}

It is clear, this scalar at $r=r_{s}$, besides $r=0$, tends to
infinity. It means that the Schwarzschild radius, as $c$ varies,
is also an essential singularity. Hence, no coordinate
transformation can be found to let the matter enter the
horizon\footnote[2]{Note that $r=r_s$ is still the radius of
hypersurfaces of $r$ equal to a constant that is the event
horizon, i.e., $g^{ab}\frac{\partial r}{\partial x^a}\frac{\partial
r}{\partial x^b}=0\quad\Rightarrow\quad r=r_s$.}, i.e. nothing can
pass the horizon. The interior part of a black hole is separated
from its exterior, i.e., the whole manifold. This situation remains
the same until the speed of light reaches a constant value during an
epoch of time. Considering the Schwarzschild radius as an
essential singularity in the VSL models can also be found in the
literature\cite{r19}, however, in another context.

The mass $M$ in (\ref{2.3}) is not only the interior mass of a
black hole, but any outer matter that is attracted towards a
black hole (especially in a spherically symmetric configuration)
should be included in $M$. Hence by attracting \footnote[3]{We use
the word ``attracting'' here instead of ``absorbing'' in the
classical models, to clear up any possible misunderstandings of
the essential singularity properties of the Schwarzschild radius
derived above.} more matter, the mass $M$, and, respectively the
Schwarzschild radius, which is the only parameter describing a
black hole, increases. Although this is not due to the VSL theory, it
guaranties the dynamic evolution of the Schwarzschild radius in
the VSL theory.

To see the effect of the VSL theory on the Schwarzschild radius, from (\ref{2.3}), one
can write
\begin{equation}
\frac{\dot r_s}{r_s}=-2\frac{\dot c(t)}{c(t)}\label{2.5}\ \ ,
\end{equation}
for a black hole with fixed mass $M$. Regarding (\ref{2.5}), when
$c$ is decreasing, i.e., $\dot{c}(t)<0$, the Schwarzschild radius
increases with time. In the case of $\dot{c}(t)>0$, the
Schwarzschild radius decreases.

The above effects are removed when the speed of light takes a
fixed value. The Lorentz-invariance is then restored and matter can
pass the Schwarzschild horizon of the black hole.

In the early Universe, when the speed of light was much higher
than now (at least $10^{30}$ higher in order to solve the SBB
problems~\cite{r6}), (\ref{2.3}) shows that the
Schwarzschild radius associated with a particular mass $M$, was
much smaller, therefore it should have caused the probability
of creation of primordial black holes to decrease. This means that
the population of primordial black holes is much lower than expected,
hence, the probability of any possible observation of these black
holes could seriously tend to zero.

\section{The charged black holes}

In this section, we extend the above procedure for the charged
black holes in the case of the VSL theory. Hence, regarding the same
analogy for the Reissner-Nordstr\o m metric, one may assume the
following solution
\begin{equation}
ds^2=\left(1-\frac{2GM}{c(t)^2r}+\frac{\varepsilon^2}{r^2}\right)c(t)^2dt^2-\left(1-\frac{2GM}{c(t)^2r}+\frac{\varepsilon^2}{r^2}\right)^{-1}dr^2-r^2d\Omega^2\label{3.1}\
\ ,
\end{equation}
where $\varepsilon$ is interpreted as the net (geometric) electric charge of
the black hole. Therefore, the Riemann tensor scalar invariant
will be
\begin{eqnarray}
R_{abcd}R^{abcd}&=&12\frac{(\frac{r_s}{r}-\frac{\varepsilon^2}{r^2})^2
+\frac{3}{4}(\frac{\varepsilon^2}{r^2})^2}{r^4}-8\frac{(\frac{r_s}{r})\ddot{c}(t)(\frac{r_s}{r}-3\frac{\varepsilon^2}{r^2})}{r^2c(t)^3(1-\frac{r_s}{r}+\frac{\varepsilon^2}{r^2})^2}{}\nonumber\\
&&{}+16\frac{(\frac{r_s}{r})\dot{c}(t)^2[(1+\frac{\varepsilon^2}{r^2})(\frac{r_s}{r}-6\frac{\varepsilon^2}{r^2})+(\frac{r_s}{r})^2]}{r^2c(t)^4(1-\frac{r_s}{r}+\frac{\varepsilon^2}{r^2})^3}{}\nonumber\\
&&+4\frac{(\frac{r_s}{r})^2\ddot{c}(t)^2}{c(t)^6(1-\frac{r_s}{r}+\frac{\varepsilon^2}{r^2})^4}-32\frac{(\frac{r_s}{r})^2\dot{c}(t)^2\ddot{c}(t)(1+\frac{\varepsilon^2}{r^2})}{c(t)^7(1-\frac{r_s}{r}+\frac{\varepsilon^2}{r^2})^5}{}\nonumber\\
&&{}+64\frac{(\frac{r_s}{r})^2\dot{c}(t)^4(1+\frac{\varepsilon^2}{r^2})^2}{c(t)^8(1-\frac{r_s}{r}+\frac{\varepsilon^2}{r^2})^6}
\label{3.2}\ \ .
\end{eqnarray}
where $\varepsilon$ is assumed to be a constant. Again, as is
obvious, the above scalar in the case of
$\varepsilon^2\leq(GM/c^2)\equiv m^2$, at
$r=r_\pm$, where
\begin{equation}
r_\pm=m\pm (m^2-\varepsilon^2)^\frac{1}{2}\label{3.3}\ \ ,
\end{equation}
also tends to infinity, hence it leads to essential singularities
there. However, in the case of $\varepsilon^2>m^2$, there is only
the usual essential singularity at $r=0$.

It is well worth noting that by assuming $\varepsilon=const.$, one
has forced the usual electric charge to vary, so we have
\begin{equation}
\varepsilon=\frac{1}{c(t)^2}\left(\frac{G}{4\pi\epsilon_0}\right)^\frac{1}{2}q\label{3.4}\
\ ,
\end{equation}
where $\epsilon_0=\frac{1}{\mu_0c^2}$. Now, by a plausible
assumption that $\mu_0$ is a constant, the electric charge should
vary as $q\propto \frac{1}{c}$, to keep $\varepsilon$ as
a fixed value. Any other assumption about the dependency or constancy
of the electric charge results in more complicated calculations. For example, in the case of $q=const.$, an analogous
metric would be
\begin{equation}
ds^2=\left(1-\frac{2GM}{c(t)^2r}+\frac{\left(\frac{\mu_0G}{4\pi}\right)q^2}{c(t)^2r^2}\right)c(t)^2dt^2
-\left(1-\frac{2GM}{c(t)^2r}+\frac{\left(\frac{\mu_0G}{4\pi}\right)q^2}{c(t)^2r^2}\right)^{-1}dr^2
-r^2d\Omega^2\label{3.5}\ \ .
\end{equation}

\section{Conclusions and remarks}

We have used a vacuum case for the VSL theory, i.e., $T_{ab}=0$ where
$G_{ab}\neq 0$. With this assumption we have achieved the following.
\begin{itemize}
    \item In spite of time-dependence, the generalized Schwarzschild
          metric~(\ref{2.7}) remains spherically symmetric. Hence,
          with $c(t)$, and not $c(t,r)$, the metric can only radiate
          with monopole symmetry.
    \item The Schwarzschild radius of a (non charged) black hole
          is not a removable singularity. No matter can pass through
          the horizon.
    \item The Schwarzschild radius is dynamic even though no matter
          is attracted towards the black hole. It increases when
          $c$ is decreasing and vice versa.
    \item If, in the very early epoch of the Universe, the value of
          $c$ was much more larger than its current
          value, then the probability of creation of the primordial black
          holes, hence their population, and consequently the
          probability of observing them is greatly reduced.
    \item When one considers a charged black hole with a fixed
          $\varepsilon$, essential singularities, besides $r=0$,
          arise only if $\varepsilon^2\leq m^2$ at $r=r_\pm$.
\end{itemize}

It should be emphasized that if $c$ attains a fixed value, then all
the above effects are switched off and the classical properties of
the black holes are restored. Actually, this is the result of the way
we embarked on the generalization of the Schwarzschild metric. However, one
should aware that simply writing in a time variation in the
Schwarzschild solution may not lead to an exact behavior, for example, in
the Brans-Dicke theory of gravitation, when $G$ is time dependent,
\emph{other} solutions arise including vacuum Friedmann Universes
that are not black holes. Although, once again, as there is not
yet a concrete dynamical theory for the VSL models, we aim to
consider their consequence(s) through a self-consistent procedure
in this work. Nevertheless, when varying-$c$ creates extra terms
in the Einstein equation, it should alter the previous dynamics,
as, usually a change in the dynamics is mainly achieved by altering
the Lagrangian of the system. For example, one knows that adding higher
derivative correction terms to a Lagrangian does not only mean
that they will perturb the original  theory, but their presence,
as unconstrained terms even with small coefficients, make the new
theory completely different from the original one~\cite{r28}.
However, having an extended horizon for black holes looks somehow
consistent with a change of speed of light.

Finally, while we were revising this manuscript, we came across
ref.~\cite{r24} in which a similar conclusion for horizon
growth is deduced.

\section*{Acknowledgement}
HS thanks his supervisor MF. This work has been
supported by the Shahid Beheshti University.

\end{document}